\documentclass[12pt]{JHEP3}
\usepackage{epsfig}
\usepackage{amsmath}
\title{
The Shape of Instantons : 
Cross-Section of Supertubes and Dyonic Instantons}
\author{
Heng-Yu Chen$^{*a}$, Minoru Eto$^{\dagger b}$, 
and Koji Hashimoto$^{\dagger c}$\\
${}^*$ {\it DAMTP, Centre for Mathematical Sciences, \\ 
\hspace{5mm}University of Cambridge, Wilberforce Road,
Cambridge CB3 0WA, UK}\\
${}^\dagger$ {\it Institute of Physics, the University of Tokyo, Komaba,
Tokyo 153-8902, Japan}\\
$^a$ E-mail: \email{h.y.chen@damtp.cam.ac.uk}\\
$^b$ E-mail: \email{meto@hep1.c.u-tokyo.ac.jp}\\
$^c$ E-mail: \email{koji@hep1.c.u-tokyo.ac.jp}\\
}

\abstract{
We explore the correspondence between Yang-Mills instantons and
algebraic curves. The curve is defined by Higgs zero locus of dyonic
instantons in 1+4 dimensional Yang-Mills-Higgs theory, and it is
identified in string theory with the cross-section of supertubes
connecting parallel D4-branes. To present evidence for the
identification, we show that with total charges fixed, the supertube
angular momentum computed from the Higgs zero locus is maximized when
the locus is circular, which has been proven for the cross-section of
the supertubes. This leads to a consistent dictionary between the
charges in two pictures. We also consider a T-dual version of the story,
identifying the profiles of the wavy instanton strings with those of the 
supercurves/D-helices. Based on this observation, we then argue a novel
correspondence between ADHM data of instantons and algebraic curves
defining the locus. The degree of the curve is related to the instanton
number, and splitting property of the curve is physically manifested by
well-separated instantons.
}

\preprint{
{\normalsize{\tt hep-th/0609142}}\\
{\normalsize DAMTP-2006-60}\\
{\normalsize UT-Komaba/06-7}
}

\begin{document}

\section{Introduction}
\label{section1}

Given a Yang-Mills instanton configuration in terms of gauge fields
in Euclidean 4 dimensions, 
how can one extract concrete physical information of
``geometry'' of the instanton --- location, size, relative moduli, and
so on? A basic strategy for this problem is to introduce a probe and
study how the probe looks at the instanton as a background geometry.
For example, introduction of a fundamental massless 
matter fermion provides us with possible Dirac zero modes. In fact,
these modes are responsible for reconstructing the ADHM data
\cite{Atiyah:1978ri} of the instantons, which is called inverse ADHM
construction \cite{Corrigan:1983sv}.  
Another probe which we consider in this paper is a Higgs field $\phi$ 
in an adjoint representation, which experiences the self-dual instanton
as a background. When $\phi$ acquires a vacuum expectation value 
(at the spatial infinity), there is a back-reaction to the instanton
configurations. However, if one adds a 
time direction by embedding the Euclidean Yang-Mills theory into 1+4
dimensions, then one can turn on an 
electric field $E_\mu \equiv F_{0\mu}$ 
to retain the instanton configuration intact: this is a
supersymmetric {\it dyonic instanton} found by 
N.~Lambert and D.~Tong \cite{Tong}, 
whose 1/4 BPS equations are defined by
\begin{eqnarray}
 F_{\mu\nu}=*F_{\mu\nu}, \quad D_\mu \phi = E_\mu,\quad 
 D_0 \phi = 0,
\label{BPSeq}
\end{eqnarray}
with $\mu,\nu = 1,2,3,4$. The solutions to these equations should also
satisfy   
the Gauss's law $D_\mu E_\mu = ie\left[\phi,D_0\phi\right]$.
The time-independent solution $(\partial_0 =0)$ is then obtained by
$A_0 = -\phi$ where $\phi$ is determined by the Laplace equation
\begin{eqnarray}
 D_\mu D_\mu \phi = 0
\label{Laplace}
\end{eqnarray}
in the instanton background \cite{Tong}.
The dyonic instantons carry electric charges as well as their
instanton charges, as a result
they have angular momenta via ``Poynting vectors''. 
The Higgs probe field $\phi$ manifests this interesting structure in its
zero locus $\phi=0$. The plot of the Higgs zero in the two instantons 
background
with arbitrary electric charge \cite{Kim} forms a
closed loop rather than just a collection of points. Higgs zero locus
can be regarded as the location of the instantons\footnote{In \cite{Kim}
the locus is interpreted as a monopole string.} 
(as in the case of 
't Hooft-Polyakov monopoles), hence this directly indicates that
the instantons have angular momenta and are ``running'' along the loop.

\begin{figure}[t]
\begin{center}
\begin{minipage}{7cm}
\begin{center}
\includegraphics[width=7cm]{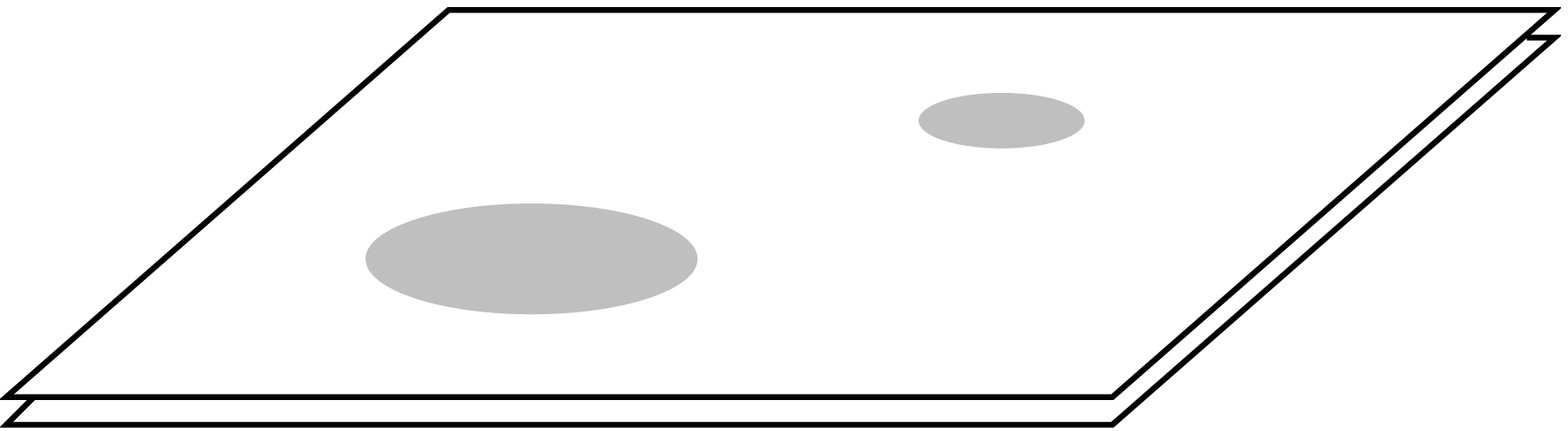}
\caption{D-brane representation of the Yang-Mills instantons. 
The D0-branes (resolved by condensation of D0-D4 strings) are
sitting inside the parallel coincident D4-branes.}
\label{fig-a}
\end{center}
\end{minipage}
\hspace{5mm}
\begin{minipage}{7cm}
\begin{center}
\includegraphics[width=7cm]{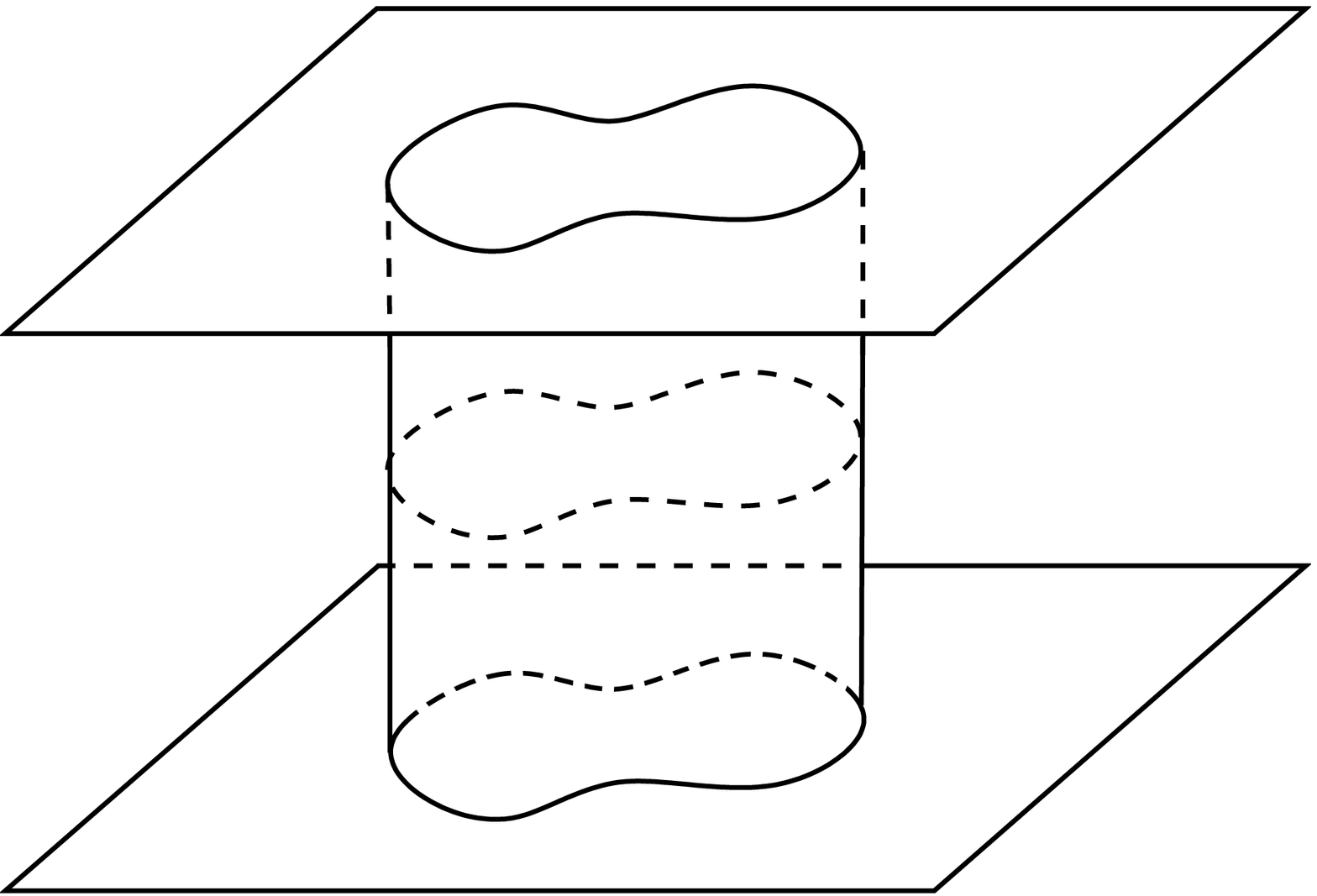}
\caption{A dyonic instanton. The D2-brane supertube is suspended between 
 two parallel D4-branes.}
\label{fig-b}
\end{center}
\end{minipage}
\end{center}
\end{figure}

Among fruitful interplay between field theory solitons and D-branes in
string theory, Yang-Mills instantons have played the role of 
touchstones. The ADHM data of the instantons 
has an interesting interpretation as the excitations on D0-branes
sitting inside D4-branes \cite{Witten:1995gx}\footnote{
Recently, the inverse ADHM construction
described above is shown to have physical interpretation in terms of
D-brane anti-D-brane annihilation \cite{Hashimoto:2005qh}.} 
(see Fig.~\ref{fig-a}). 
The D-brane interpretation of the dyonic instantons is a supertube
\cite{Mateos:2001qs,Emparan:2001ux,Mateos:2001pi} 
connecting parallel (but separated) D4-branes
(Fig.~\ref{fig-b}) \cite{Kruczenski:2002mn}.\footnote{Some discussions
on D-brane interpretation of the dyonic instantons are present in
\cite{Lambert:1999ix}.}  
The charges, masses and supersymmetries of the dyonic instantons 
can be identified consistently with their D-brane counterparts. 
The supertube consists of a tubular D2-brane 
on which fundamental strings (F1)
and D0-branes are bound, and from the perspective of D4-branes, 
they appear as a monopole string, the electric charges and 
the usual instantons respectively. The supertube preserves 1/4 of the
supersymmetries maintained by the D4-branes, while the dyonic instantons 
preserve 1/4 supersymmetries of the 1+4 dimensional super Yang-Mills
theory. 

The supertubes have arbitrary cross-sections while keeping 
their stability and supersymmetries, this is due to the fact that the 
D0-branes running along the D2-brane surface keeps the
shape of the cross-section against collapsing to a point.
In other words, the tension of the tubular D2-brane is canceled out
by the angular momentum, so the supertube is stable without collapsing.

At this stage, 
the Higgs zero locus of the
dyonic instantons may naturally be identified with 
the cross-section of the supertubes
\cite{Kim}.
To provide further evidence for such identification, it is natural to
ask whether the supertubes continue to exhibit its intrinsic properties
with additional inputs from its proposed field counterparts, the dyonic
instantons. 
For example, it is well known that the angular momentum of the supertube
is given by the curve defining its cross-section \cite{Mateos:2001pi}.  
In section \ref{section2} of this paper, we would like to show that, by
substituting the Higgs zero locus for the cross-section in the
expression for supertube angular momentum, {\it{such angular momentum is
maximized at circular Higgs zero locus, provided the electric charge
$Q_{e}$ of the dyonic instanton is kept fixed}}. This specific property
for supertube was first demonstrated in the supergravity contexts
\cite{Mateos:2001pi}, here we consider a purely field theoretical
calculations and formulate the analogous variational problem.     
This in turns would require us to set up a dictionary between the
conserved charges of dyonic instantons and 
supertubes. We shall argue that this dictionary is crucial in giving
consistent interpretation between our field theory results and the known 
stringy properties of the supertubes. 
In addition, we study a T-dual picture of this story in section
\ref{supercurve}, demonstrating that the momentum for the so called
``D-helix'' \cite{d-helix} (or its S-dual -- ``supercurve''
\cite{Mateos:2002yf}) has identical functional dependence on its
shape as their field theory counterparts known as ``wavy
instanton strings'' \cite{EINO}. This identification in fact is where
such a dictionary is at its clearest.

The identification of the Higgs zero locus with the supertube
cross-section 
leads us to an interesting correspondence: {\it real algebraic curves}
$\leftrightarrow$ {\it Yang-Mills instantons}. The Higgs zero of the
dyonic instantons serves as a bridge between these two concepts.
Note that the Higgs field does not spoil the original instanton 
configuration, because the second equation in (\ref{BPSeq})
does not cause a back-reaction to the first (self-dual) equation.
In this sense the Higgs field is a true probe, with a help of the 
electric field. The Higgs zero loci are given by algebraic real curves 
which are defined as polynomial equations of spatial coordinates.
In section \ref{section3}, we examine this intriguing correspondence.
We shall see that the degree of the algebraic curves is related to the
instanton number, and the splitting property of the curves is manifested 
physically by well-separated instantons.

\section{Maximization of angular momentum}
\setcounter{footnote}{0}
\label{section2}

\subsection{Dyonic instantons and angular momentum}
\label{review}

The dyonic instantons are BPS solutions of ${\cal N}=2$ super
Yang-Mills(-Higgs) theory\footnote{
The dyonic instantons are also 1/4 BPS objects in theories with 8
supercharges.}  
in 1+4 dimensions \cite{Tong}. In view of the
correspondence to the D-branes, we concentrate on SU(2) gauge group
provided by the two parallel D4-branes for simplicity. 
The Higgs zero loci can form non-trivial loops when we have more than
one instanton, 
and they appear as the monopole strings in the 1+4 dimensions.
A complete treatment of arbitrary instanton numbers would require
solving the intricate full ADHM constraints. However in this section we
shall focus on the case of two instantons where one can bypass such
difficulties, and we shall extensively utilize the results obtained  
by S.~Kim and K.~Lee \cite{Kim}. They explicitly solved the equations
(\ref{BPSeq}) using the ADHM method inspired by 
the Jackiw-Nohl-Rebbi (JNR) ansatz \cite{Jackiw:1976fs}, 
\begin{eqnarray}
 A_\mu(x) = \frac{i}{2}\sigma^a \bar{\eta}_{\mu\nu}^a \partial_\nu
\log
H_{\rm JNR}(x),\quad
H_{\rm JNR}(x) \equiv
\sum_{i=0}^2 \frac{s_i}{|y_i|^2} \ ,
\label{jnr}
\end{eqnarray}
where $\bar\eta^a_{\mu\nu}$ is the anti-self-dual 't Hooft tensor
and $y_{i\mu} \equiv x_\mu -a_{i\mu}$.
The parameters specifying the instantons are
the three scalar moduli $s_{0},s_{1},s_{2}$ which are real and positive,
while the three 
position moduli are labeled as $a_{0\mu},a_{1\mu}, a_{2\mu}$, where
$\mu=1,\dots,4$ labeling the four spatial directions. Note that for two
instantons, this ansatz has enough number of the degrees of freedom to
sweep the entire instanton moduli space. Without losing generality, we 
can rotate the configuration and set $a_{i1}=a_{i2}=0$ ($i=0,1,2$). 
Other moduli of the dyonic instanton are the spatially asymptotic values 
$q^a \sigma^a$ of the Higgs field $\phi$. For simplicity we consider the
probe Higgs field with $q^1=q^2=0$ which gives the Higgs zero on the 
3-4 plane and was considered in \cite{Kim} in details:
\begin{eqnarray}
\phi(x)\big|_{y_{i1}=y_{i2}=0} = 
\frac{q^3\sigma_3}{(s_0+s_1+s_2) H_{\rm JNR}(x)}
\frac{X(x)}{|y_0|^2|y_{1}|^2|y_2|^2}\ .
\label{phi}
\end{eqnarray}
Here we have defined a polynomial function
\begin{eqnarray}
X(x) \equiv
\sum_{i=0}^2 \left(\!
s_i^2 |y_{i+1}|^2 + 2 s_is_{i+1}y_{i}\! \cdot\! y_{i+1}
- 
\frac{4 \left(y_{i}\! \times\! y_{i+1}\right)
\sum_{j=0}^2(a_j\!\times\! a_{j+1})}
{\sum_{j=0}^2 (s_j s_{j+1})^{-1}|a_j\!-\!a_{j+1}|^2}
\right) |y_{i+2}|^2\ , \;\;
\label{X}
\end{eqnarray}
where $\vec{y}_i \equiv \vec{x}-\vec{a}_i$ 
are two dimensional vectors on the 3-4 plane, and the cross product 
is defined as $y_i \times y_{j}\equiv y_{i3}y_{j4}-y_{i4}y_{j3}$.

Explicit solution (\ref{phi}) shows \cite{Kim} that the Higgs
zero locus is given by the following real algebraic curve
\begin{eqnarray}
X(x) = 0.\label{X=0}
\end{eqnarray}
The electric charge of the dyonic instanton is given as
\begin{eqnarray}
 Q_{\rm e} &=& {\rm tr}\int\! d^4x \; \partial_\mu (E_\mu \phi)  
\nonumber \\
&=&\frac{8\pi^2(q^3)^2}{(s_0\!+\!s_1\!+\!s_2)^2}
\left\{
\sum_{i=0}^2s_is_{i+1}|a_i-a_{i+1}|^2
-
\frac{4 \sum_{i=0}^2(a_i\times a_{i+1})^2}
{\sum_{j=0}^2 (s_j s_{j+1})^{-1}|a_j-a_{j+1}|^2}
\right\} \ . \quad \label{Qelectric}
\end{eqnarray}
One can obviously see that both $X(x)=0$ and $Q_{\rm e}$ are 
invariant under the overall scaling of the scalar moduli $s_{i}$, 
which are defined projectively.

The identification between the supertube in string theory and dyonic
instantons in field theory naturally leads one to further identify the
Higgs zero locus $X=0$ with the curve defining supertube
cross-section. The ideal place to test such further identification is to
consider the angular momentum.   
According to \cite{Mateos:2001pi}, 
the angular momentum of the supertube is given by 
\begin{eqnarray}
 L = 
\oint_{-\pi}^\pi \! d\theta
\left(
x_3 \frac{\partial x_4}{\partial \theta}-
x_4 \frac{\partial x_3}{\partial \theta}
\right) \, ,
\label{angL}
\end{eqnarray}
where $\theta$ is the coordinate on the curve defining the cross-section.
Suppose we made such further identification between the curves in field
theory and string theory. 
Once the ADHM data $(s_i, a_{i\mu})$ is provided for the dyonic
instanton, one can then compute $X=0$ and obtain the curve, and
substitute this 
curve profile into (\ref{angL}) to compute the angular momentum for the
supertube. The idea here is that we pose a variational problem in field
theory which extremizes (\ref{angL}), with the dyonic instanton charges
fixed. 
This is related to the similar variational problem for the supertube in
string theory, with the supertube charges fixed \cite{Mateos:2002yf}.  
Even though we are varying essentially the same integral (\ref{angL}),
however these two problems are not a priori identical, as the charges
which we kept fixed in the field theory and string theory can be
different functionals of the curves. 
To show these two problems can be identified, one has to propose some
kind of dictionary which relates the charges in these two theories, and
we shall explain such dictionary further in the next subsection
\ref{dictionary}.  
Conversely we shall see in section \ref{maxl} that, for fixed electric
charge $Q_{e}$ of the 
dyonic instanton, the angular momentum (\ref{angL}) calculated from the
Higgs zero locus is maximized when it forms a circle, this provides
strong evidence for such dictionary between the charges.  

To make the distinctions clear, one should first note that the
definition (\ref{angL}) for the supertube only captures the angular
momentum along the Higgs zero locus given by (\ref{X=0}), and this
generally differs from the usual field theoretical definition for the
angular momentum of 
dyonic instanton, which is given by a four dimensional integral
\cite{Kim,Eyras:2000dg},\footnote{See discussions in section
\ref{section4}.}    
\begin{eqnarray}
 \widetilde{L}_{\mu\nu}
= \int \! d^4x \left(x_\mu T_{0\nu}-x_\nu T_{0\mu}\right) 
\label{AngST}
\end{eqnarray}
with $T_{\mu\nu}$ being the energy momentum tensor
of the 1+4 dimensional Yang-Mills theory.

In the following, 
we shall first explain the proposed dictionary between charges in more
details. We then give the support for such dictionary by demonstrating
that for fixed $Q_{e}$, the angular 
momentum for the supertube as defined by (\ref{angL}) is maximized when
the Higgs zero locus forms a circle. 
To do so, we derive in section \ref{condcircle} 
a condition for $X=0$ to give a circle in terms of the
ADHM data of the instantons.
In section \ref{maxl}, we show that perturbations 
around the ADHM data giving the circle always decrease the angular
momentum. There we also present some numerical calculations.
In addition, we consider a more evident example of the dictionary
between 
charges given in \ref{supercurve},  
where D-helices/supercurves \cite{d-helix,Mateos:2002yf} (embedded in
two coincident D5-branes) are identified with wavy instanton string
solutions \cite{EINO} in 1+5 dimensional SU(2) Yang-Mills theory. For
this case, the dictionary between the charges is completely proven.

\subsection{Dictionary between supertubes and dyonic instantons}
\label{dictionary}
\setcounter{footnote}{0}

Here we would like to discuss further the dictionary mentioned earlier
between the conserved charges for the dyonic instantons and the
supertubes. 
On one hand if we had assumed such dictionary, then variational problem
in field theory proposed earlier would be identical to the one for the
supertubes in string theory.  
On the other hand, our results in section \ref{maxl}, which demonstrates
that the angular momentum (\ref{angL}) calculated from the Higgs zero
locus maximizes at the circle for fixed dyonic instanton charges, can   
be regarded as the supporting evidence for such dictionary. 

First let us list the standard relations between 
the various parameters in string theory and the 1+4 dimensional
SU(2) Yang-Mills-Higgs theory. The distance between the
D4-branes, $l$, is 
related to the Higgs expectation value at asymptotic infinity,
\begin{eqnarray}
l = 2\pi l_s^2 \cdot 2q^3
\label{defdis}
\end{eqnarray}
with $l_{s}$ being the string length. 
The gauge coupling constant $e$ in the Yang-Mills-Higgs theory is 
related to the string coupling constant through the tension of the
D4-brane as $1/e^2 = (1/2) T_{\rm D4}(2\pi l_s^2)^2$
which can be simplified as 
\begin{eqnarray}
e^2 = 8 \pi^2 l_s g_s \, . 
\label{defcoup}
\end{eqnarray}

Consider the BPS equations (\ref{BPSeq}), it describes a dyonic
instanton whose energy is given in terms of its charges \cite{Tong}:   
\begin{eqnarray}
 {\cal E}_{\rm dyonic}= \frac{8\pi^2}{e^2} \kappa + \frac{2}{e^2} 
Q_{\rm e} \ . 
\label{enedi}
\end{eqnarray}
Here $\kappa$ is the instanton number, and  
$Q_{\rm e}$ is the electric charge of the dyonic instanton given 
in equation (\ref{Qelectric}).
This energy should be equal to that of the supertube. 
The supertube is described as a configuration of D2-brane worldvolume
theory, and the energy is just given by the tension of the supertube
(per unit length) 
times its length $l$ (which equals the distance between the two
D4-branes),
\begin{eqnarray}
 {\cal E}_{\rm supertube} = l \int_{-\pi}^{\pi}\!\! d\theta\; {\cal H}. 
\end{eqnarray}
Here $\theta$ is the cylindrical coordinate parameterizing the
cross-section of the supertube surface, 
and ${\cal H}$ is the energy density of the supertube given
in terms of a low energy D2-brane worldvolume theory.
Using the decomposition of the Hamiltonian by the D-brane/F-string
charges given in \cite{Emparan:2001ux}\footnote{We use the notation
given in \cite{Bak}. The quantization of the charges is given by
$2\pi Q_{\rm F1}\in {\bf Z}$ and $l Q_{\rm D0}\in {\bf Z}$.}, 
we arrive at the expression 
\begin{eqnarray}
 {\cal E}_{\rm supertube} = l \left(
T_{\rm D0} Q_{\rm D0}+2\pi T_{\rm F1} Q_{\rm F1}
\right) \ .
\label{enesu}
\end{eqnarray}
Here $T_{\rm F1}$ $(Q_{\rm F1})$ and $T_{\rm D0}$ $(Q_{\rm D0})$ are 
the tension
(charge) of the F-strings and the D0-branes respectively.
Comparing (\ref{enedi}) and (\ref{enesu}), and using the standard
relations (\ref{defdis}) and (\ref{defcoup}), we propose the dictionary
\begin{eqnarray}
&&\frac{\kappa}{q^3} = 4\pi l_{\rm s}^2 Q_{\rm D0} \ , 
\label{instn}
\\
&&\frac{Q_{\rm e}}{q^3} = 16\pi^3 l_{\rm s} g_{\rm s}  Q_{\rm F1} \ .
\label{instn2}
\end{eqnarray}
Suppose the identifications (\ref{instn}) and (\ref{instn2}) 
are valid, they would imply
that fixing $\kappa$ and $Q_{e}$ for dyonic instantons corresponds
precisely to keeping fixed the D0 and F1 charges for the supertube,
hence the two variational problems can be identified and the maximal
angular momentum occurs at the circular shape. Conversely, we will show
in section \ref{maxl} that angular momentum for the supertube
(\ref{angL}) 
calculated from the Higgs zero locus is maximized when it becomes
circular, for given $\kappa$ and $Q_{e}$, this should then be 
interpreted as a supporting evidence for the dictionary (\ref{instn})
and (\ref{instn2}). 
 
The identification (\ref{instn}) and (\ref{instn2})
means not only that the values of the charges are identical,
but also that the charges are identical functionals of the 
curves which are defined by the Higgs zero locus for the dyonic
instantons and by the cross-section for the supertubes.
This is the main point of the present paper, so here  
we explain in details the logic of our reasoning given briefly above. 

The supertubes are described by the
following fields on the D2-brane: the magnetic field $B(\theta)$, 
the conjugate momentum $\Pi_z(\theta)$ with respect to the gauge field, 
and the transverse scalar field $y_\mu(\theta)$. The supertubes satisfy
the relation 
\begin{eqnarray}
B\Pi_z= T_{\rm D2}|y_\mu'(\theta)|^2.  
\label{supereq}
\end{eqnarray}
The charges are defined as
$Q_{\rm D0}= (1/2\pi)\int\! d\theta \;B$, 
$Q_{\rm F1}= (1/2\pi)\int\! d\theta \;\Pi_z$.
Using reparameterization of $\theta$, we may put the magnetic field 
constant. 
Through the equation (\ref{supereq}), the electric charge
is dependent on the
shape of the cross-section $y_\mu(\theta)$. Given the cross-section
$y_\mu$, we can define the angular 
momentum of the supertube $L_{\rm supertube}$ as in (\ref{angL}), thus
the angular momentum 
is a functional of the cross-section,  
$L_{\rm supertube}[y_\mu]$.
On the other hand, for the dyonic instantons, the instanton number
$\kappa$ is fixed while the electric charge $Q_{\rm e}$ is a function 
of the ADHM
data $(s_i,a_i)$. The Higgs zero locus is defined by the ADHM data, so
the curve $x_\mu = \hat x_\mu (\theta)$ is given 
as a solution of $X(s_i, a_i, x)=0$. The angular momentum $L$ 
is defined by this $\hat x_\mu$, as explained in section \ref{review},
so $L$ is 
a functional of $\hat x_\mu$, $L=L[\hat x_\mu]$. Since the definition is
the same as the 
supertube, we should note that the $\hat x_\mu$ dependence in $L$ is the
same as  
the $y_\mu$ dependence in $L_{\rm supertube}$. 

From these precise definitions, the meaning of the ``dictionary''
(\ref{instn}) and (\ref{instn2})
is clear. The left hand side of (\ref{instn2}) is a
function of $(s_i,a_i)$ while the right hand side
is a functional of $y_\mu$. Note that 
we can extract the information of the ADHM data $(s_i,a_i)$ from the 
curve $\hat x_\mu $, so implicitly $Q_{\rm e}$ is a functional 
of the curve $\hat x_\mu$. The proposed dictionary means that,
the functional dependence of the both sides are the
same, if we identify the curve $\hat x_\mu$ with the cross-section
$y_\mu$. 
The result of section \ref{maxl} will strongly support that this is
correct, since the variational problems in the two pictures exhibit 
the same property:
the angular momentum is maximized when the curve is circular.
This in turn means that the identification of the 
Higgs zero locus with the supertube cross-section, $\hat{x}_\mu=y_\mu$,
is correct. 

\subsection{Conditions for circular cross section}
\label{condcircle}

In this and next subsections, we shall show that the maximization of the 
supertube angular momentum (\ref{angL}) 
computed from the Higgs zero locus $X=0$
is achieved when the locus is circular. First, let us study for which
value of the ADHM data the curve $X=0$ becomes circular.

For the circular cross-section to appear,
i.e. $(x_{3}^{2}+x_{4}^{2})=r_0^{2}$, it is necessary for $X$ to take
the following form 
\begin{equation}
A(x_{3}^{2}+x_{4}^{2})^{2}+B(x_{3}^{2}+x_{4}^{2})+C=0
\label{circularcond} \, ,
\end{equation}
that is a quadratic equation in $(x_{3}^{2}+x_{4}^{2})$, the radius of
the circle $r_0$ is then given by the root of this equation. 
Here we show that the Higgs zero locus 
(\ref{X=0}) forms a circular shape (\ref{circularcond}) with
a unique set of the nine parameters $(s_i, a_{i\mu})$ ($i=0,1,2$ and
$\mu=3,4$), up to the overall scaling of $s_i$ and the rotation around
the origin $x_3=x_4=0$. 

Consider the most general fourth order
polynomial for $x_{3}$ and $x_{4}$,  
\begin{eqnarray}
&&b_{4}x_{3}^{4}+b_{3}x_{3}^{3}+b_{2}x_{3}^{2}+b_{1}x_{3}
+c_{4}x_{4}^{4}+c_{3}x_{4}^{3}+c_{2}x_{4}^{2}+c_{1}x_{4}+k\nonumber\\
&&+d_{1}x_{3}x_{4}+d_{2}x_{3}^{2}x_{4}^{2}+f_{1}x_{3}x_{4}^{2}
+f_{2}x_{3}^{2}x_{4}+h_{1}x_{3}x_{4}^{3}+h_{2}x_{3}^{3}x_{4}
\label{general4th}\,.
\end{eqnarray}    
Clearly by comparing (\ref{X}) with (\ref{circularcond}) and
(\ref{general4th}), we seem to have more conditions than the number of
variables. However, $X$ as
defined in (\ref{X}) belongs to a special subset of these general
polynomials, simple expansion can show that for $X$, the followings are
satisfied identically:
\begin{eqnarray}
b_{4}=c_{4}=\frac{1}{2}d_{2}=(s_{0}+s_{1}+s_{2})^{2}\,,\quad
b_{3}=f_{1}\,,\quad c_{3}=f_{2}\,,\quad
h_{1}=h_{2}=0\,.
\end{eqnarray}
We are left with precisely six conditions
\begin{equation}
b_{3}=0\,,~c_{3}=0\,,~d_{1}=0\,,~b_{2}=c_{2}\,,~b_{1}=0\,,~c_{1}
=0\,.\nonumber
\end{equation}
The first four conditions give rise to the following
equations for $a_{i3}$ and $a_{i4}$: 
\begin{eqnarray}
&&a_{03}(s_{1}+s_{2})+a_{13}(s_{0}+s_{2})+a_{23}(s_{0}+s_{1})=0\,,
\label{con1}\\
&&a_{04}(s_{1}+s_{2})+a_{14}(s_{0}+s_{2})+a_{24}(s_{0}+s_{1})=0\,,
\label{con2}\\
&&s_{0}(a_{13}a_{24}+a_{14}a_{23})+s_{1}(a_{03}a_{24}+a_{04}a_{23})
+s_{2}(a_{03}a_{14}+a_{04}a_{13})=0\,,
\label{con3}\\
&&s_{0}(a_{13}a_{23}-a_{14}a_{24})+s_{1}(a_{03}a_{23}-a_{04}a_{24})
+s_{2}(a_{03}a_{13}-a_{04}a_{14})=0\,.
\label{con4}
\end{eqnarray}
The first two equations are solved straightforwardly as 
\begin{eqnarray}
 a_{03} = \frac{-\left(
a_{23}(s_0\! +\! s_1) + a_{13}(s_0\! +\! s_2)
\right)}{(s_1 + s_2)} \,, \quad 
 a_{04} = \frac{-\left(
a_{24}(s_0\! +\! s_1) + a_{14}(s_0\! +\! s_2)
\right)}{(s_1 + s_2)} \,.\label{a03a04}
\end{eqnarray}
Then (\ref{con3}) is solved as 
\begin{eqnarray}
 a_{13} = \frac{-a_{23}s_1(a_{24}(s_0 + s_1) + a_{14}s_2)}
{s_2(a_{24}s_1 + a_{14}(s_0 + s_2))}, 
\label{a1324}
\end{eqnarray}
where we have implicitly assumed that the denominator is non-vanishing.
Using these we solve (\ref{con4}) as
\begin{eqnarray}
 a_{23} = \pm \sqrt{\frac{s_2}{s_0s_1(s_0 + s_1 + s_2)}}
(a_{24}s_1 + a_{14}(s_0+s_2)).
\label{pm}
\end{eqnarray}
We choose the negative sign in the right hand side of (\ref{pm}) for
simplicity. The condition $b_1=0$ is then simplified to
\begin{eqnarray}
 (s_0 + s_2)(s_0 + s_1 -2s_2)a_{14}
= (s_0 + s_1)(s_0 + s_2 -2s_1)a_{24}.
\label{1424}
\end{eqnarray}
It is straightforward to show\footnote{To show this, the positivity of
$s_i$ is crucial. If any one of them becomes zero, the instanton number
reduces by one. And we exclude the possibility that the circle collapses
to a point.} 
that this condition is compatible with
the last equation $c_1=0$ with (\ref{a03a04})-(\ref{pm}) only 
if $s_0=s_1=s_2$.
All the constraints are solved with this as 
\begin{eqnarray}
 a_{03}=-a_{13}-a_{23}\,, \;
a_{04} = -a_{24}-a_{14}\,, \;
a_{13} = -a_{23}
\frac{a_{14}\!+\! 2a_{24}}{a_{24}\! +\! 2a_{14}}\,,
\;
a_{23} = \frac{-(a_{24}\!+\!2a_{14})}{\sqrt{3}}
\,.
\nonumber
\end{eqnarray}
Substituting these into $X$, we find
\begin{eqnarray}
 \frac{X}{(s_0)^2}=
9(x_3^2\!+\!x_4^2)^2 + 4 (a_{14}^2\! +\! a_{14}a_{24}\! +\! a_{24}^2)
(x_3^2\!+\!x_4^2) -\frac{16}{3}(a_{14}^2\! +\! a_{14}a_{24} 
\!+\! a_{24}^2)^2
\hspace{10mm}
\label{X_kimlee}
\end{eqnarray}
which certainly gives a single circle.\footnote{
If we choose the positive sign in
(\ref{pm}), the reasoning is found to be the same except that
we have parity flipped constraints, $a_{i3} \leftrightarrow -a_{i3}$.
} 
The example of the ADHM data
given in \cite{Kim}, 
\begin{eqnarray}
&&s_0^{(0)}=s_1^{(0)}=s_2^{(0)}=1\,, \;
(a_{03}^{(0)}, a_{04}^{(0)}) = (-R, 0)\,,\; \label{cm}\\
&&(a_{13}^{(0)}, a_{14}^{(0)}) = (R/2,-\sqrt{3}R/2)\,,\;
(a_{23}^{(0)}, a_{24}^{(0)}) = (R/2, \sqrt{3}R/2),\nonumber
\end{eqnarray}
is a particular representative of our general solution.\footnote{Note
that the overall scaling of the parameters $s_i$ doesn't change
anything, that is why we could fix the overall normalization of $s_i$
in (\ref{cm}).}
The example is the unique data giving the circle,
up to rotation around the origin and parallel transport on the 3-4
plane.

\subsection{Maximization around circular profile}
\label{maxl}

Let us study the perturbations from the circular Higgs zero found above
and see how the angular momentum (\ref{angL}) is maximized there.
It is enough to consider the representative (\ref{cm}) according to the
uniqueness shown in the previous subsection. The radius of the circle is
given by 
\begin{eqnarray}
 r_0 = \sqrt{\frac{\sqrt{13}-1}{6}}R,\label{r0}
\end{eqnarray}
thus the angular momentum for the configuration (\ref{cm})
is given by 
\begin{eqnarray}
 L = 2\pi r_0^2 = 
2\pi \frac{\sqrt{13}-1}{6}R^2 = 
\frac{\sqrt{13}-1}{48\pi} \frac{Q_{\rm e}}{(q^3)^2}.
\label{ange}
\end{eqnarray}
In the last equation we have used the electric charge for this ADHM
data, $Q_{\rm e}=16\pi^2 (q^3)^2 R^2$.

\subsubsection{Showing the extremum}

The small perturbations from the point (\ref{cm}) are given by 
\begin{eqnarray}
 s_i \equiv 1 + \delta s_i\,,\quad
 a_{i3} \equiv a_{i3}^{(0)} + R\delta a_{i3}\,,\quad
 a_{i4} \equiv a_{i4}^{(0)} + R\delta a_{i4}\,.
\label{perturbations}
\end{eqnarray}
Let us consider the first order in perturbation. 
We expand the function $X$ around the data (\ref{cm}).
A straightforward calculation gives 
\begin{eqnarray}
 X = \hat{X} 
+ X_i \delta s_i + R X_{i3} \delta a_{i3}
+ RX_{i4} \delta a_{i4} + {\cal O}(\delta^2),
\end{eqnarray}
where $\hat{X}$ is the polynomial given in equation
(\ref{X_kimlee}) and  $X_{i}, X_{i3}$ and $X_{i4}$ are homogeneous
polynomials 
of $(x_3, x_4, R)$, whose explicit expressions can be obtained easily
(we don't show them here).

In solving $X=0$, we may use the following parameterization for the
coordinates $x_3$ and $x_4$:
\begin{eqnarray}
 x_3 \equiv (r_0 + R\delta r(\theta)) \cos(\theta)\,,\quad
 x_4 \equiv (r_0 + R\delta r(\theta)) \sin(\theta)\,. 
\label{eq:parametrization_dr}
\end{eqnarray}
The fluctuation of ${\cal O}(\delta r)$ coming from the
substitution of this into $\hat{X}$ is then given by 
\begin{eqnarray}
 \hat{X} = 
0 + \sqrt{78(-1+\sqrt{13})}R^4 \delta r 
+ {\cal O}((\delta r)^2)\label{Xhat}\,.
\end{eqnarray}
Therefore we may solve the equation $X=0$ to derive the expression of
the deformation of the radius $\delta r$ written in terms of the moduli
deformation $\delta s$ and $\delta a$ as 
\begin{eqnarray}
 \delta r(\theta) = \frac{-
\left[
X_i \delta s_i + RX_{i3} \delta a_{i3}
+ RX_{i4} \delta a_{i4} 
\right]_{x_3 = r_0\cos\theta,\; x_4 = r_0 \sin\theta}}
{\sqrt{78(-1+\sqrt{13})}R^4}\label{deltar}\,.
\end{eqnarray}
With this, we can compute the angular momentum 
along the Higgs zero locus as\footnote{
It is interesting to note that at this order the
deformation $\delta s_i$ does not appear, somehow.
}
\begin{eqnarray}
 L &=& 
\int \! d\theta (r_0^2 + 2 r_0 R\delta r + {\cal O}((\delta r)^2))
\nonumber \\
&\simeq&
2\pi r_0^2 - \frac{\pi}{3}(\sqrt{13}-1) R^2
\left[
 \frac{-2}{3}\delta a_{03}
+ \frac{1}{3}\delta a_{13}
+ \frac{1}{\sqrt{3}}\delta a_{14}
+ \frac{1}{3}\delta a_{23}
+ \frac{1}{\sqrt{3}}\delta a_{24}
\right]\,.
\hspace{10mm}\label{pertL}
\end{eqnarray}

We need to
evaluate how the electric charge $Q_{\rm e}$ responds to the
perturbation. In 
other words, keeping $Q_{\rm e}$ fixed imposes a
linear constraint among the deformation parameters.
A straightforward calculation then gives
\begin{eqnarray}
 \frac{Q_{\rm e}}{(q^3)^2 16\pi^2 R^2}  = 
1
+ \frac{-2}{3}\delta a_{03}
+ \frac{1}{3}\delta a_{13}
+ \frac{1}{\sqrt{3}}\delta a_{14}
+ \frac{1}{3}\delta a_{23}
+ \frac{1}{\sqrt{3}}\delta a_{24}
+ {\cal O}(\delta^2)
\,,\nonumber
\end{eqnarray}
thus fixing the electric charge is equivalent to the constraint
\begin{eqnarray}
 \delta a_{03}= \frac12 \left[
\delta a_{13}
+ \sqrt{3}\delta a_{14}
+ \delta a_{23}
+ \sqrt{3}\delta a_{24}
\right]\,.\label{LinearCon}
\end{eqnarray} 
Now it is obvious that the perturbation of $Q_{\rm e}$ coincides with
the perturbation of $L$ (\ref{pertL}). This means that, if we fix the
electric charge 
$Q_{\rm e}$, then the angular momentum does not change at the first
order perturbation. 
We showed that, for the angular momentum,
(\ref{cm}) is a stationary point in the moduli space.
This is a strong evidence that the angular momentum
is maximized at the circular configuration (\ref{cm}).
We shall demonstrate next that this is indeed a local maximum, by
considering the perturbation at second order.

\subsubsection{Showing the maximum}

We have nine moduli parameters $(s_i,a_{i3},a_{i4})$ in general, 
these are too many variables to analyze exactly
at the second order in their perturbations. Thus, for simplicity,
we restrict ourselves to treat a subspace of the moduli space,  
and consider a more manageable situation with only four parameters 
$\delta a_{03},\ \delta a_{13},\ \delta a_{23}$ and $\delta a_{04}$
while fixing all the others to the values (\ref{cm}), 
to demonstrate the maximization of the angular momentum. 

The variation of the electric charge $Q_{\rm e}$ with respect to
this perturbation is then 
\begin{eqnarray}
\frac{Q_{\rm e}}{16\pi^2 (q^3)^2R^2} &=& 
1 
+ \frac{1}{3}
\left(- 2\delta a_{03}\! +\! \delta a_{13}\! +\! \delta a_{23} \right) 
+ \frac{4}{9}\left( (\delta a_{03})^2\! + (\delta a_{13})^2 
\!+ (\delta a_{23})^2\! + (\delta a_{04})^2 \right) 
\nonumber\\
&& \hspace{-15mm}
- \frac{2}{9}\left(
\delta a_{03} \delta a_{13} + \delta a_{13} \delta a_{23} 
+ \delta a_{23} \delta a_{03} 
- \sqrt{3} \delta a_{13} \delta a_{04} 
+ \sqrt{3} \delta a_{23} \delta a_{04} \right)
+ {\cal O}(\delta^3)\,.
\nonumber
\end{eqnarray}
We want to fix the value of the electric charge 
$Q_{\rm e} = 16\pi^2 R^2 (q^3)^2$ which is given by (\ref{cm}).  
This leads to a constraint
which can be solved for $\delta a_{23}$ up to this order, as\footnote{
Note that the first order constraint 
$\delta a_{13} + \delta a_{23} = 2\delta a_{03}$ 
is as itself inconsistent with the second order part of the constraint 
$- 4 (\delta a_{03})^2
- 4 (\delta a_{13})^2 - \frac{4}{3} (\delta a_{04})^2
+ 8 \delta a_{03}\delta a_{13} 
+ \frac{4}{\sqrt 3} \delta a_{03} \delta a_{04}
- \frac{4}{\sqrt 3} \delta a_{13} \delta a_{04} = 0$.
One should not solve the constraint 
order by order.
} 
\begin{eqnarray}
\delta a_{23} &=&
\left[
2\delta a_{03} - \delta a_{13}
\right]
+ 
\left[
 - 4 (\delta a_{03})^2
- 4 (\delta a_{13})^2 - \frac{4}{3} (\delta a_{04})^2 
\right.\nonumber\\
&&
\hspace{20mm}
\left.
+\  8 \delta a_{03}\delta a_{13} 
+ \frac{4}{\sqrt 3} \delta a_{03} \delta a_{04}
- \frac{4}{\sqrt 3} \delta a_{13} \delta a_{04}
\right]
+ {\cal O}(\delta^3)\,.
\label{eq:2nd_a13_a03}
\end{eqnarray}

Let us next solve the condition $X=0$ by using 
the parameterization (\ref{eq:parametrization_dr}).
Plugging the perturbed variables into the equation $X=0$,
we can express $\delta r(\theta)$ in terms of 
$\delta a_{03},\ \delta a_{13},\ \delta a_{04}$ 
and  $\delta a_{23}$, up to their second orders. 
The explicit expressions are not so illuminating,
however the resultant 
angular momentum up to the second order takes a very simple form
\begin{eqnarray}
L = 2\pi r_0^2 - 2\pi R^2 \alpha
\left(- \sqrt 3 \delta a_{03} + \sqrt 3 \delta a_{13} 
+ \delta a_{04}\right)^2\!\!,
\;
\alpha \equiv \frac{137 \sqrt{13}\! -\! 13^2}{13^2\cdot 3^2}
(\approx 0.21)\,.\hspace{10mm}
\label{eq:L1}
\end{eqnarray}
Here we have eliminated $\delta a_{23}$ by 
using (\ref{eq:2nd_a13_a03}).
As expected, the first order terms are vanishing and 
the coefficient of the second order is negative, 
$ - 2\pi R^2 \alpha$.
This is the corroboration of our claim 
that (\ref{cm}) gives the unique maximum of the angular momentum.

It is not so easy to increase the number of perturbation parameters,
thus we analyze here some other sets of parameters to illustrate the
point. 
For the set 
$(\delta a_{03},\ \delta a_{23},\ \delta a_{04},\ \delta a_{14})$, 
the same procedures give the angular momentum 
\begin{eqnarray}
L = 2\pi r_0^2 - 2\pi R^2 \alpha
\left[
\left( - \sqrt 3 \delta a_{03} + \delta a_{04} 
- 2 \delta a_{14} \right)^2
+ 3 \left( \delta a_{14} \right)^2
\right].
\label{eq:L2}
\end{eqnarray}
For the set 
$(\delta a_{03},\ \delta a_{23},\ \delta a_{14},\ \delta a_{24})$, 
we obtain 
\begin{eqnarray}
L = 2\pi r_0^2 - 2\pi R^2 \alpha
\left[
\left( - \sqrt 3 \delta a_{03} - 2 \delta a_{14} 
+ \delta a_{24}\right)^2
+ 3 \left( \delta a_{14} - \delta a_{24} \right)^2
\right].
\label{eq:L3}
\end{eqnarray}
For the set
$(\delta a_{23},\ \delta a_{04},\ \delta a_{14},\ \delta a_{24})$, 
the angular momentum is of the form
\begin{eqnarray}
L = 2\pi r_0^2 - 2\pi R^2 \alpha
\left[
\left( \delta a_{04} - 2 \delta a_{14} + \delta a_{24}\right)^2
+ 3 \left( \delta a_{14} - \delta a_{24} \right)^2
\right].
\label{eq:L4}
\end{eqnarray}
In all cases, the angular momentum (\ref{angL}) is maximized at the
circular Higgs  
zero locus (\ref{cm}).\footnote{
Note that the above four results (\ref{eq:L1}), (\ref{eq:L2}), 
(\ref{eq:L3}) and (\ref{eq:L4}) are consistent with each other.
A naive conjecture for the angular momentum with the perturbation
of six parameters $\delta a_{i3}$ and $\delta a_{i4}$ around the 
circular profile is then given by
\begin{eqnarray*}
L = 2\pi r_0^2 - 2\pi R^2 \alpha
\left[
\left( -\sqrt 3 \delta a_{03} + \sqrt 3 \delta a_{13} 
+ \delta a_{04} - 2 \delta a_{14} + \delta a_{24}\right)^2
+ 3 \left( \delta a_{14} - \delta a_{24} \right)^2
\right]
\end{eqnarray*}
where we have eliminated $\delta a_{02}$ by using the condition 
to keep $Q_{\rm e}$ fixed.
}

\subsubsection{Numerical evaluation}
\setcounter{footnote}{0}
Since the analytical evaluation is intricate, we turn to
a numerical evaluation of the angular momentum (\ref{angL}) for given
ADHM data of the instantons. We find that, even if we include
perturbation of the 
parameters $s_i$ in addition to $a_{i\mu}$, the angular momentum is
still maximized at the circular profile (\ref{cm}). 

\begin{figure}[t]
\begin{center}
\begin{minipage}{3.5cm}
\begin{center}
\includegraphics[width=3.5cm]{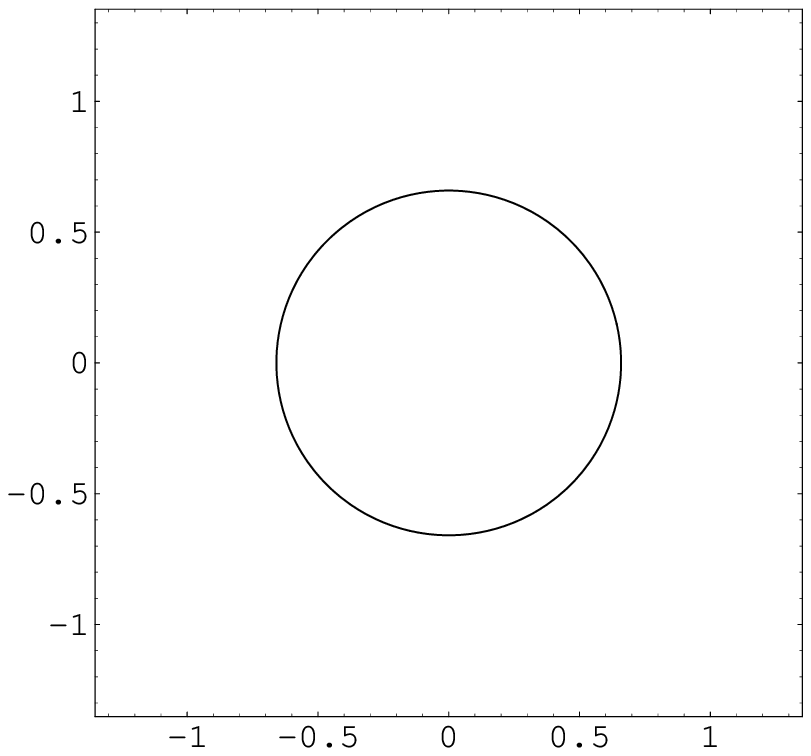}
\caption{$s_0=1$.}
\label{fig1}
\end{center}
\end{minipage}
\begin{minipage}{3.5cm}
\begin{center}
\includegraphics[width=3.5cm]{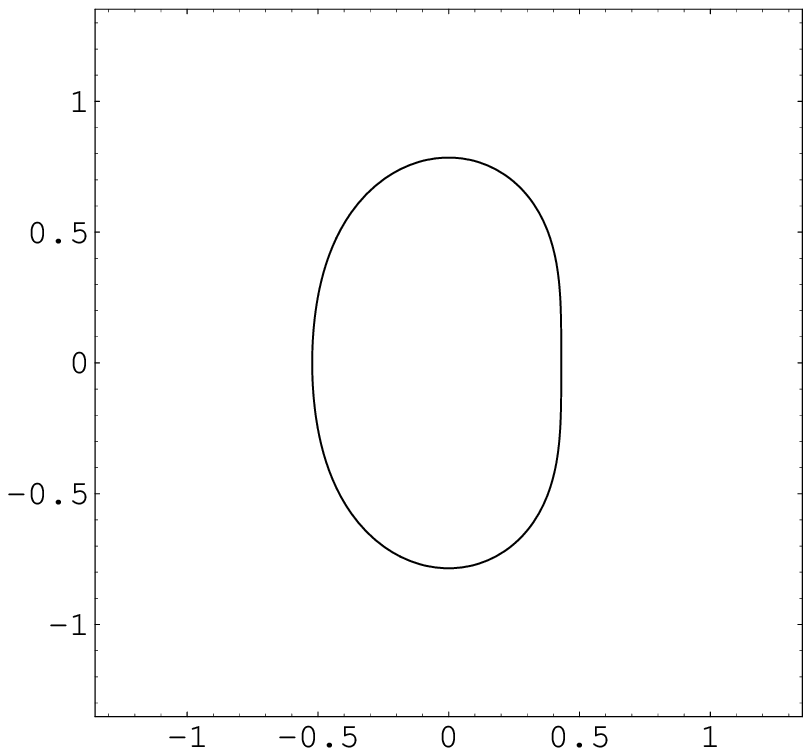}
\caption{$s_0=4$.}
\label{fig2}
\end{center}
\end{minipage}
\begin{minipage}{3.5cm}
\begin{center}
\includegraphics[width=3.5cm]{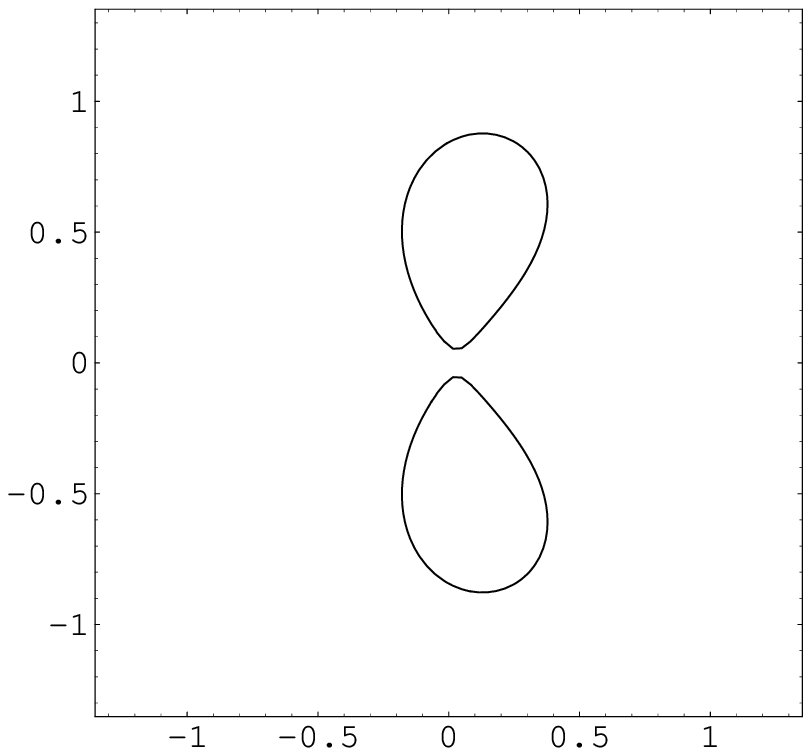}
\caption{$s_0=13$.}
\label{fig3}
\end{center}
\end{minipage}
\begin{minipage}{3.5cm}
\begin{center}
\includegraphics[width=3.5cm]{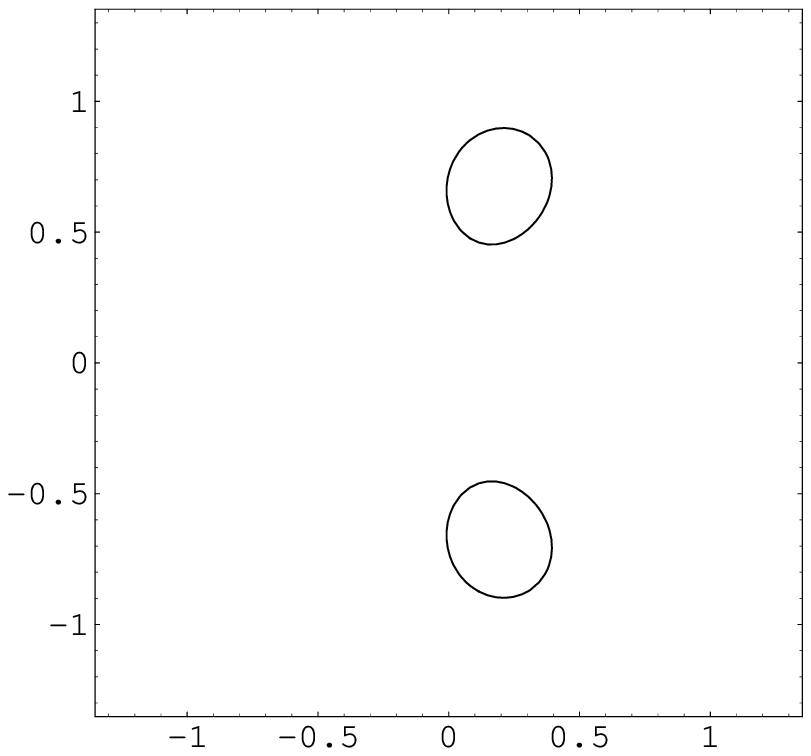}
\caption{$s_0=25$.}
\label{fig4}
\end{center}
\end{minipage}
\end{center}
\end{figure}

To illustrate the numerical simulations, here we present some peculiar
examples of the deformation to the circular profile. The most
interesting property of the loops given by the Higgs zero loci is that 
they split as one changes the ADHM data. See Fig.~\ref{fig1}, 
$\cdots$, Fig.~\ref{fig4}. We have chosen $a_{03} = -\sqrt{s_0}R$ 
and varied $s_0$, while other variables are kept to be (\ref{cm}).
As before, we keep the electric charge fixed, which determines  the
parameter $R$. For $s_0=1$ which is (\ref{cm}), we have a circle,
but as we increase the parameter $s_0$, the circle is deformed
and splits into two loops, and the loops continuously shrink to
vanishing size at large $s_0$. The large $s_0$ limit is identical to the
't Hooft instanton, 
as seen in (\ref{jnr}) and discussed in \cite{Kim}.

The angular momenta (\ref{angL}) calculated from the Higgs zero loci are
summarized in the following table. One can 
immediately see that it is maximized at $s_0=1$
of (\ref{cm}).\footnote{
The angular momentum is proportional to the area
enclosed by the loops, as seen from the definition (\ref{angL}).
One can observe that the area is in fact decreasing as one increases
the parameter $s_0$.
} The cases with $s_0<1$ show the maximization at $s_0=1$,
too, though we have not listed the evaluated numerical values.
The angular momenta in the table are measured in the unit of 
$Q_{\rm e}/(16\pi^2 (q^3)^2)$.
\begin{center}
\begin{tabular}{|c||c|c|}
\hline
 &$s_0$ & angular momentum \\
\hline
\hline
Fig.~\ref{fig1} & 1 & 2.73 $(=2\pi(\sqrt{13}-1)/6)$ \\
\hline
Fig.~\ref{fig2} & 4 & 2.47 \\
\hline
Fig.~\ref{fig3} & 13 & 1.38 \\
\hline
Fig.~\ref{fig4} & 25 & 0.57\\
\hline
\end{tabular}
\end{center}

\subsection{Wavy instanton strings and supercurves/D-helices}
\label{supercurve}

To illustrate the dictionary given in section \ref{dictionary}, 
here we study the IIB counterpart of the supertube and the corresponding 
BPS soliton in Yang-Mills theory.\footnote{Related calculations are 
found in \cite{Redi}.}
We shall see that, interestingly, in this case the dictionary can be
proven as a functional of the curves.

When we take a T-duality along the axial direction of the supertube, 
we find a super D-helix which is a helical D1-brane moving with 
the speed of light along its axis \cite{d-helix}.
As an example, let us consider a perfectly circular D-helix with 
radius $R$ for definiteness
(the radius is defined as a trajectory projected on a plane
perpendicular to the axis), which is 
the T-dual of a perfectly circular supertube with the same radius.
The Hamiltonian of the super D-helix is given by
${\cal H} = T_{\rm D1}|B| + |\Pi|$, where $\Pi$ is the momentum of the
D-helix along the axis, 
and $B$ is the slope of the D-string along the axis,
namely the pitch of the D-helix is given by $2\pi R B$. 
The relation to the supertube is apparent: in view of (\ref{enesu}),
the momentum corresponds to the F1 charge of the
supertube, and the slope corresponds to the D0-brane charge.
The super D-helix satisfies 
$\sqrt{2\pi l_{\rm s}^4 g_{\rm s}|\Pi B|} = R$, which means
that the speed of the D-helix along the axis is equal to the speed of
light. By the T-duality, this is equivalent to the equation 
$E=\pm 1/2\pi l_{\rm s}^2$
of the supertube.
Moreover, the super D-helix
preserves a quarter of the bulk supersymmetries, which is the same as
the supertube. 

Supercurves formulated in \cite{Mateos:2002yf}
are the objects that are S-dual to the D-helices, 
and there it was found that the
supercurves, which are fundamental strings traveling with the speed of
light, allow 
arbitrary deformation of the shape as long as the speed is maintained,
which is analogous to the supertubes. 
The Hamiltonian is similar to the above,
\begin{eqnarray}
{\cal H} = T_{\rm F1}|Z'(\theta)| + |P_Z(\theta)|\,.
\label{hamilcurve}
\end{eqnarray}
Here $Z(\theta)$ 
is a scalar field on the worldsheet and specifies the location
along the axis of the supercurve, and 
we can choose the gauge $Z'=1$ by using reparameterization of the
worldsheet coordinate\footnote{In this subsection we regard $\theta$ as
a dimensionful parameter, for our convenience. 
One can realize this by a redefinition 
$\theta \to \theta l_{\rm s}$.} $\theta$. $P_Z$ is the momentum of the
supercurve 
along the axis $Z$. The shape of the supercurve projected onto the plane
transverse to the $Z$ direction is given by the other scalar fields 
$y^\mu(\theta)$. The BPS equations of motion shows 
\begin{eqnarray}
P_Z(\theta) = T_{\rm F1}|y_\mu'(\theta)|^2 \, .
\label{momentumsuper}
\end{eqnarray}
As long as this relation is satisfied, 
$y_\mu$ dependence in the Hamiltonian vanishes.
This is the reason why the supercurves allow almost 
arbitrary deformations. 

It was further shown in \cite{Mateos:2002yf} that, for fixed total
momentum $\int\! d\theta P_Z$ of the periodic supercurves, 
the angular momentum
defined by $y_\mu(\theta)$ as in (\ref{angL}) is maximized when the
projected shape $y_\mu$ of the supercurve is circular. This is
consistent with the T-duality.
Since we can apply the same method for the
D-helices, in the following we shall use the notation of the
supercurves. 

In order to find a field theoretical counterpart of the super D-helix,
let us consider a T-duality of the situation at hand: a supertube 
suspended between two parallel D4-branes as we considered for the 
dyonic instantons in this paper.
Taking the T-duality along the transverse direction of the 
D4-branes, we obtain
the super D-helix embedded in coincident two D5-branes.
As the supertubes suspended between the D4-branes
can be thought of as the 1/4 BPS dyonic instantons in 1+4 dimensions
from the D4-brane view point, the super D-helix should be thought of 
as a 1/4 BPS soliton in a 1+5 dimensional Yang-Mills theory, 
from the D5-brane view point.
Such a 1/4 BPS soliton in 1+5 dimensions was found 
in \cite{EINO} as a solution of the 1/4 BPS equations
\begin{eqnarray}
F_{\mu\nu} = * F_{\mu\nu},\quad
E_\mu = F_{5\mu},\quad 
E_5 = 0,
\label{wavyeq}
\end{eqnarray}
with the Gauss's law $D_\mu E_\mu + D_5 E_5 = 0$ ($\mu,\nu = 1,2,3,4$).
The solution is called {\it wavy instanton strings}, which extends along 
the $x^5$ axis 
on which an electric wave runs with the speed of light. The $x^5$
direction is identified with the $Z$ direction of the 
D-helices/supercurves. 

Notice that all the BPS equations for the wavy instanton string 
(including the Gauss's law) and their solutions
lead to those for the dyonic instantons
by the dimensional reduction $\partial_5 = 0$ and $A_5 = - \phi$ 
\cite{EINO}.
The dimensional reduction can be naively thought of as embodiment of the 
T-duality in field theories, analogous to the well-known dimensional
reduction from instantons (D4-D0 system) to monopoles (D3-D1 system). 

The BPS equations (\ref{wavyeq}) can be solved by choosing
$\partial_0 = \partial_5$ and $A_0 = A_5$ in any instanton background 
\cite{EINO}. Namely, concrete solutions of the wavy instanton string 
can be found by promoting the moduli parameters $y^\mu$ for the center 
of the instantons to be arbitrary functions of the combination $x^0+x^5$
as  
\cite{EINO}
\begin{eqnarray}
A_\mu = A_\mu^{\rm inst}\left(x^\mu - y^\mu(x^0 + x^5)\right)\,, 
\quad
A_0=A_5=-\frac{\partial y^\mu}{\partial x^0} A_\mu \, .
\label{wavysol}
\end{eqnarray}
For example, a helical wavy instanton string with circular profile on
the projected plane can be obtained by choosing the particular functions
\begin{eqnarray}
y^3 = R \cos \left(\frac{x^0 + x^5}{ RB}\right),\quad
y^4 = R \sin \left(\frac{x^0 + x^5}{ RB}\right),
\end{eqnarray}
while $y^1,y^2$ are constants. This helical wavy instanton string has
the radius $R$, its velocity is the speed of right and its pitch is
$2\pi R B$, the same as the ``circular'' super D-helix described above.

Obviously, the correspondence between the super D-helix and helical wavy
instanton string is more transparent than that between the supertubes
and the dyonic instantons. Here, we shall directly show that the
momentum 
computed in the field theory language has identical functional 
dependence on $y^\mu$ as that of the supercurve $P_Z$. 
This is a proof of the dictionary analogous to the one proposed in the
case of the dyonic instantons $\leftrightarrow$ the supertubes. 

The Hamiltonian of the wavy instanton strings is given by 
\begin{eqnarray}
 {\cal H} = 
\int\! dx^5
\left[
\frac{8\pi^2}{e^2}\kappa + \frac{2}{e^2}
\int\! d^4x \; {\rm tr} F_{0\mu}F_{5\mu}
\right]\, .
\label{wavyham}
\end{eqnarray}
One can see the complete analogy with the dyonic instanton,
(\ref{enedi}). The second term is the momentum along the $x^5$ axis.
In comparison to the supercurves/D-helices, the Hamiltonian
(\ref{hamilcurve}) shows that the dictionary is
\begin{eqnarray}
\frac{8\pi^2}{e^2}\kappa = T_{\rm D1} |Z'|\, , \quad 
\frac{2}{e^2}
\int\! d^4x \; {\rm tr} F_{0\mu}F_{5\mu}
= |P_Z|\, . 
\label{dicwavy}
\end{eqnarray}
The first equation should be understood in the gauge $Z'=1$ and with
$\kappa=1$, since we consider a single D-helix/supercurve.

The Yang-Mills field strength for the single instanton is given by 
\begin{eqnarray}
 F^{\rm inst}_{\mu\nu} = \frac{\rho^2}{(|x_\mu|^2+\rho^2)^2}
\bar{\eta}_{\mu\nu}^{a}\sigma_a.
\label{bpst}
\end{eqnarray}
According to the solution (\ref{wavysol}), the location of the center of 
the wavy instanton string is just given by $x^\mu=y^\mu(x^0+x^5)$. This
is almost clear from the definition, but one can define this location in
a gauge invariant manner as a solution to the equations 
${\rm tr}\partial_\mu (F_{\rho\sigma}F_{\rho\sigma})=0$, for
example.\footnote{For the dyonic instantons studied in the previous
subsections, this identification has not been clear, and that is why we
studied the angular momentum problem in details. Here the identification
is clear.} With this single instanton solution, let us evaluate the
momentum given by the second term of (\ref{wavyham}). 
From (\ref{wavysol}), it is
straightforward to obtain 
\begin{eqnarray}
 F_{0\mu} = F_{5\mu}=
-\frac{\partial y^\rho}{\partial x^5}F_{\rho\mu}\, .
\end{eqnarray}
Therefore, 
\begin{eqnarray}
 \int\! d^4x \; {\rm tr} F_{0\mu}F_{5\mu} = 
\frac{\partial y^\rho}{\partial x^5}
\frac{\partial y^\sigma}{\partial x^5}
\int\! d^4x \; {\rm tr} F^{\rm inst}_{\rho\mu}
F^{\rm inst}_{\sigma\mu}.
\end{eqnarray}
Here we have made a shift $x^\mu-y^\mu(x^0+x^5)\to x^\mu$ in the 
integral since it does not affect the result.
The integral can be evaluated with the BPST instanton
(\ref{bpst}), but we just need the dependence on the indices, 
\begin{eqnarray}
\int\! d^4x \; {\rm tr} F^{\rm inst}_{\rho\mu}
F^{\rm inst}_{\sigma\mu}
=({\rm const.}) \times \bar{\eta}_{\rho\mu}^a
\bar{\eta}_{\sigma\mu}^a
=({\rm const.}) \times \delta_{\rho\sigma}\, .
\end{eqnarray} 
Thus we find 
\begin{eqnarray}
\int\! d^4x \; {\rm tr} F_{0\mu}F_{5\mu} = 
({\rm const.}) \times 
\biggl|
\frac{\partial y^\rho}{\partial x^5}
\biggr|^2\, .
\end{eqnarray}
This is precisely what we have expected in (\ref{momentumsuper}),
in the gauge $dx_5/d\theta = Z'=1$.
The dictionary (\ref{dicwavy}) is proven as a functional of the shape of
the wavy instanton strings and the supercurves/D-helices.

The virtue of this T-dual example, compared to the dyonic instantons,
is that from the first place, the wavy instanton string solution
(\ref{wavysol}) includes the functional dependence explicitly. 
However, the dimension reduction to dyonic instantons is not easy,
and so far there are only a limited number of solutions (such as periodic
instantons) found to be consistent with the explicit T-duality in string
theory picture.  
It would be interesting if the T-duality in field theories can be
formulated more usefully, to provide a route for proving the dictionary
of the dyonic instantons, from that of the wavy instanton strings.

\section{Algebraic curves capturing ADHM data}
\setcounter{footnote}{0}
\label{section3}

\subsection{A conjecture}
\label{section3-1}

The central concept underlying the analysis in the previous section
is that the ADHM data, once it is given, produces a real algebraic
curve. This is very intriguing in the sense that ADHM data are generally
difficult to interpret, especially when the number of instantons is
large.  Another difficulty is due to the presence of the gauge orbit 
among the ADHM data. In view of this, the algebraic real curves (loops)
given by the Higgs zero may be a good and effective alternative
to capture the information of the instantons.\footnote{Related
references along a similar spirit can be found in
\cite{Stovicek:1989hu}.} 

\begin{figure}[t]
\begin{center}
\includegraphics[width=10cm]{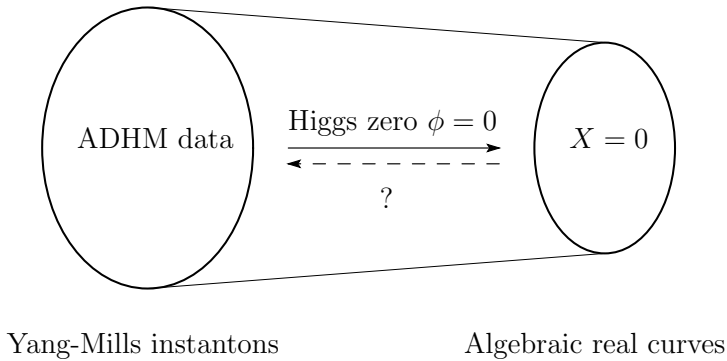}
\caption{Correspondence between ADHM data
and curves by Higgs zero.}
\label{loop_adhm}
\end{center}
\end{figure}

One may hope that the correspondence is one-to-one and thus the 
curves may capture {\it all} the information of the instantons.
(For monopoles, certainly this kind of correspondence exists:
spectral curves of monopoles \cite{spectral}, 
which are Riemann surfaces.)
We know that for a given instanton configuration, it defines
a Higgs zero, thus the map is at least on-to, for appropriately
defined set of algebraic curves (recall that coefficients of the
polynomial are related, as seen in section \ref{condcircle}). 
In general, the requirement of adjoint $\phi=0$ consists of three real
equations, while we have four spatial coordinates $x_\mu$, thus
resulting in a one dimensional curve.
The concept is depicted in Fig.~\ref{loop_adhm}.

Unfortunately, one can find some counter-examples which shows that
the correspondence is not one-to-one. 
Let us first consider the circle. 
It is expected that the circular Higgs zero locus appears
for any instanton number, there appears to be degeneracy to one
another. 
However, this degeneracy can be lifted if one varies the
instanton numbers while keeping fixed all the other parameters 
involved\footnote{
This can be explicitly seen in the picture of the supertubes, since
for fixed F1 charges, the D0-brane charge per a unit length along the
supertube isometry (which, if multiplied by the length between the
D4-branes, is the instanton number) is proportional to the
enclosed area of the supertube cross-section. See section \ref{section4}
for the detailed correspondence to the D-brane pictures.
}: 
the electric charge $Q_{\rm e}$ and the asymptotic Higgs vev $q$.
We must note here that the rotation in the plane on which the circle
lies does not change the circle. But this rotation changes the instanton
profile, as seen in the previous section. One cannot lift this
rotationary degeneracy. 

Another counter-example already appears in the case of the two
instantons. Let us consider 't Hooft ansatz for the
instantons. As briefly mentioned before, the 
Higgs zero corresponding to the 't Hooft solutions is just
two points and not forming a curve \cite{Kim}. However, we know that the
't Hooft instantons have size moduli parameters $s_{i}$ in addition to
the location of the instantons. This information of the size 
is not reflected in the profile of the curve. Even when
the curve does not shrink to points, it is insensitive to
the overall scaling of $s_{i}$.

Even with these counter-examples, it is still interesting to
pursue this direction. Certainly the Higgs zero is a gauge-invariant 
object, and is simple enough because it is given by just algebraic
polynomials. Obviously, location of each instanton is indicated by the
Higgs zero, which is certainly true for the probe of the adjoint scalar 
field.

In the following, we first count the degree of the polynomial $X$ which
defines the Higgs zero, and show that the degree is bounded by the
instanton number $\kappa$. Therefore, 
when a closed curve is given in terms of 
a polynomial (as usual for algebraic curves), the degree of  
it gives the {\it minimal} number of instantons necessary for
reproducing the curve as a Higgs zero. This is analogous to the
spectral curves of monopoles \cite{spectral}, where the degree of the
polynomial defining the spectral curve 
(which is related to the genus of the corresponding Riemann surface) is
determined by monopole number. To reproduce any 
curve which is not expressed by any polynomial, we need infinite number
of instantons. The importance of this interesting limit 
$\kappa \to\infty$ has been pointed out in \cite{Townsend:2004nc}. 
Furthermore, we must note that some polynomials may
have no corresponding ADHM data. We have seen in the previous section
that some coefficients in $X$ are related to each other.

We shall show in section \ref{section3-3}
that when some of the instantons are located far away from
the rest, the curves given by $X=0$ splits into two parts. This
splitting property should be endowed with the algebraic curves,
because naively the location of the instantons is specified by the
curves. 

\subsection{Degree of algebraic curve and instanton number}
\label{section3-2}

\subsubsection{JNR instantons}

To see how the degree of the polynomial $X$ is related to the data 
of the instantons appearing in the ADHM construction, we first consider
the JNR $\kappa$ instantons.
A similar counting for the full ADHM construction will be given later.
The ADHM construction of the Higgs solution \cite{Tong,Kim,Dorey} 
for the JNR ansatz (\ref{jnr}) is given by equation (3.8) of \cite{Kim}
as  
\begin{eqnarray}
\phi (x)
= \frac{1}{s_\Sigma H_{\rm JNR}(x)} 
\left[ 
\bar Z(x) q Z(x) + Q(x)
\right]
\label{phi-jnr}
\end{eqnarray}
with $q=q^i\sigma_i$ and $s_\Sigma = \sum_{i=0}^{\kappa}s_i$.
In order for $\phi$ to be constant at spatial
infinity, the degree of 
polynomial $\bar Z(x) q Z(x) + Q(x)$ should be same as 
that of ${\rm deg}(H_{\rm JNR}) = -2$.
Indeed, the definition of $Z$ and $Q$ are given \cite{Kim} by
\begin{eqnarray}
Z(x) = \sum_{i=0}^\kappa\frac{s_iy_{i\mu}}{|y_i|^2}\ e_\mu,\quad
Q(x) = s_0^4 \Lambda^T
\left\{ \frac{1}{Y} - \frac{1}{y_0} \right\}
KpK
\left\{ \frac{1}{\bar Y} - \frac{1}{\bar y_0} \right\}
\Lambda.
\label{kpk}
\end{eqnarray}
Clearly the degrees of $\bar Z q Z$ and $Q$ are $-2$.
Here we have defined a representation of the quaternion as 
$e_\mu \equiv (i\vec\sigma,\ 1_{2\times 2})$ and 
$y_i = (x_\mu - a_{i\mu})e_\mu$. 
$p$ is a quaternionic $\kappa \times \kappa$
matrix which is determined by equation (3.3) in \cite{Kim}. 
$Y$ is a quaternionic diagonal matrix
$Y = {\rm diag}\left(y_1,y_2,\cdots,y_\kappa\right)$.
$\Lambda$ is a row vector with $\kappa$ real constant components, 
$\Lambda = \left(s_1^2/s_0^2,s_2^2/s_0^2,\cdots,s_\kappa^2/s_0^2\right)$, 
and $K$ is a constant matrix taking value in $GL(\kappa,{\bf C})$.
See \cite{Kim} for the detailed derivation. 

Note that all of $H_{\rm JNR}(x)$, $\bar Z(x) q Z(x)$ and $Q(x)$ have
second order divergence at $x_\mu = a_{i\mu}$, although $\phi(x)$ 
given in (\ref{phi-jnr}) is regular all over the space.
Now we want to derive a polynomial 
determining the zero of the adjoint field $\phi$. 
Obviously the pre-factor $1/H_{\rm
JNR}(x)$ is unrelated to the polynomial and note that $\bar Z q Z + Q$
has the negative degree as mentioned above. The negative degree can be
turned into positive by pulling out an overall factor  
$\Pi(x) \equiv (|y_0|^2 |y_1|^2 \cdots |y_{\kappa}|^2)^{-1}$
from $\bar Z q Z + Q$ as
\begin{eqnarray}
\phi (x)
= \frac{\Pi(x)}{s_\Sigma H_{\rm JNR}(x)} X_{\rm JNR}(x),\quad
X_{\rm JNR}(x) \equiv
\Pi^{-1}(x)
\left[ 
\bar Z(x) q Z(x) + Q(x)
\right].
\label{xjnr}
\end{eqnarray}
$X_{\rm JNR}(x)$ is an ordinary polynomial whose degree is
\begin{eqnarray}
{\rm deg}(X_{\rm JNR}) = 2(\kappa+1) - 2 = 2\kappa.
\label{boundjnr}
\end{eqnarray}
To be precise, the factor $\Pi^{-1}$ (which is proportional to the unity
in the quaternion) should be inserted between $Z$ and
$\bar{Z}$, and between $\left\{Y^{-1}-y_0^{-1}\right\}$ and 
$\left\{\bar{Y}^{-1}-\bar{y_0}^{-1}\right\}$ in $Q$, and in order to
cancel the divergence in those terms, we need the relation
$[y_i,q]=0$ and $[y_i,KpK]=0$. The latter equation is satisfied since
$KpK$ is the unity in the quaternion \cite{Kim}, but the former is in
general not 
satisfied. Hence we assume $q^1=q^2=0$ for simplicity, and use
the fact that putting $y_{i1}=y_{i2}=0$ consistently
solves the constraint of vanishing $\sigma_1$ and $\sigma_2$
components in $\phi$. Thus the curve is on the plane spanned by $x_3$
and $x_4$, and we end up with a single constraint 
$X_{\rm JNR}(x_3,x_4)=0$ coming from the $\sigma_3$ component of 
the equation $\phi=0$.

Note that $X_{\rm JNR}(x)$ agrees with $X(x)$ in equation (\ref{phi})
for the $\kappa=2$ instantons.\footnote{For two instantons, 
there is an interesting geometrical realization of the ADHM data
\cite{Hartshorne:1978vv}, whose relation to $X$ was discussed in
\cite{Kim}.}  
For the case of one instanton which is
always 
expressed by the 't Hooft ansatz, the locus of the Higgs zero is known to
be a point \cite{Tong} which can be written as 
$\sum_{\mu=1}^4 (x_\mu-a_\mu)^2=0$. 
This polynomial is of degree 2, ensuring the above result.

\subsubsection{Generic instantons}

For $\kappa\geq 3$, the JNR ansatz does not cover all the moduli space
of the instantons. Here we are going to show a less stringent bound 
${\rm deg}(X)\leq 4\kappa$ for general configurations  
with $\kappa$ instantons. 

The ADHM construction of the Higgs solution \cite{Tong,Kim,Dorey} 
is given, for example by equation (3.2) of \cite{Kim}, as a quaternionic
relation\footnote{Interestingly, this formula (\ref{adhmq}) has a natural
interpretation in terms of the tachyon condensation of the D-branes
\cite{Hashimoto:2005qh}: the
matrix $p$ comes from the massless excitation of a string on 
D-branes and anti-D4-branes which are to be annihilated. These are
separated from 
each other, and the location is specified by the eigenvalues of $p$.} 
\begin{eqnarray}
 \phi = v^\dagger 
\left(
\begin{array}{cc}
q & 0
 \\
0 & p
\end{array}
\right)v \, .
\label{adhmq}
\end{eqnarray}
The vector $v$ is a solution to the zero mode
equation $\Delta^\dagger v=0$
where $\Delta$ is the Dirac operator matrix whose elements are
quaternions, 
\begin{eqnarray}
 \Delta = 
\left(
\begin{array}{c}
\Lambda \\
\Omega-x1_{\kappa\times\kappa}
\end{array}
\right)
\end{eqnarray}
where a quaternion $x$ is defined as before,
$x\equiv x_1{\bf{1}} + x_2 {\bf i} + x_3 {\bf j} + x_4 {\bf k}$. 
$\Omega$ is a $\kappa\times \kappa$ matrix.
Solving the zero mode equation $\Delta^\dagger v=0$ 
formally,
\begin{eqnarray}
 v = 
\left(
\begin{array}{c}
1 \\
(\Omega^\dagger-x^*1_{\kappa\times\kappa})^{-1}\Lambda^\dagger
\end{array}
\right){\cal N}.
\end{eqnarray}
This is a $(\kappa+1)\times 1$ matrix (a vector) whose elements are
quaternions. 
The pre-factor ${\cal N}$ is the normalization of the vector
which is determined by requiring $v^\dagger v=1$ as a quaternion
equation. Using the zero modes, we obtain the Higgs field as 
\begin{eqnarray}
 \phi = {\cal N}^* 
\left(
q + \Lambda (\Omega-x1_{\kappa\times\kappa})^{-1}p
(\Omega^\dagger-x^*1_{\kappa\times\kappa})^{-1}\Lambda^\dagger
\right){\cal N}.
\end{eqnarray}
To see the Higgs zero, we may just drop the normalization 
${\cal N}$ and ${\cal N}^*$.\footnote{We 
here assumed that the normalization ${\cal N}$
does not vanish anywhere.} 
Let us evaluate the degree of $X$. For this we use 
the previous (Pauli matrix) representation of the quaternion.
Note that in this complex representation, the size of all the quaternion
matrices $\Omega$, $\Lambda$, $q$ and $p$ is doubled.
Then the equation $\phi=0$ is written in a polynomial form as 
\begin{eqnarray}
X \equiv q |\det(\Omega-x^\mu e_\mu\otimes
1_{\kappa\times\kappa})|^2 + 
\Lambda YpY^\dagger
{\Lambda}^\dagger = 0.
\end{eqnarray}
Here $Y_{ij}$ is the cofactor matrix element,
\begin{eqnarray}
 \left[(\Omega-x^\mu e_\mu\otimes
1_{\kappa\times\kappa})^{-1}\right]_{ij}
= Y_{ij}/\det(\Omega-x^\mu e_\mu\otimes
1_{\kappa\times\kappa}).
\end{eqnarray}
Using the known degrees
${\rm deg}(Y) = 2\kappa-1$, 
${\rm deg}(\det(\Omega-x^\mu e_\mu\otimes
1_{\kappa\times\kappa}))= 2\kappa$, 
it is shown that the degree of the polynomial function $X$ is
at most 
\begin{eqnarray}
 {\rm deg}(X)\leq4\kappa.
\label{generalbound}
\end{eqnarray}
This degree can be smaller than $4\kappa$, as we haven't imposed 
the ADHM equation for the ADHM data and the on-shell 
constraint for $p$. Once these constraints are
imposed, the function $X$ may be factorized and the degree
of $X$ may become less than $4\kappa$. 
In fact, the earlier JNR counting which is equivalent to the ADHM 
construction for $\kappa\le 2$ gives the degree $2\kappa$
(\ref{boundjnr}) which is below this general bound
(\ref{generalbound}).

\subsection{Splitting of the curves}
\label{section3-3}

We anticipated that the above analysis with the JNR ansatz 
exhibits the following splitting property of the algebraic curves 
for well-separated instantons.
Here we shall show that when some of the parameters $a_{i}$
are well separated from the rest, then the algebraic curve defined by
the polynomial $X$ splits into two closed curves. 

Let us divide the ADHM data $\{a_{i\mu}\}$ to two sets, 
$S^{(1)}\equiv \{a_p | \; p=0,\cdots,s\}$ and 
$S^{(2)}\equiv \{a_r | \; r=s+1,\cdots,\kappa\}$, 
and assume that two sets
are far from each other, 
\begin{eqnarray}
 |a_{p}-a_{r}|^2 \sim l\, ,\quad
 |a_{p_1}-a_{p_2}|^2 \ll l\, , \quad 
 |a_{r_1}-a_{r_2}|^2 \ll l\, ,  
\end{eqnarray}
for fixed large $l$ which is the distance between the two sets.
When $x$ is close to the first set $S^{(1)}$, 
$|y_r|$ ($r=s+1, \cdots, \kappa$) 
is very large and approximated by $l$, 
thus its dependence in $Z$ and $Q$ drops out since $y_k$ appears there
as $1/y_r \sim 1/l \to 0$. 
We arrive at an algebraic curve $X^{(1)}=0$ written only by the elements
$a_p$ of the first set $S^{(1)}$.
It gives a curve sitting around the first set $S^{(1)}$. 
Its precise expression is 
\begin{eqnarray}
 X \sim l^{2(\kappa-s)} X^{(1)}(y_p)
\label{factori}
\end{eqnarray}
where $X^{(1)}$ is of degree $2s$. 
On the other hand, when $x$ is close to the second
set, we obtain 
a curve $X^{(2)}=0$ written only by $a_r$ ($r=s+1,\cdots,\kappa$) in the
same way. 
The curve appears at the region near the second set $S^{(2)}$,
and $X \sim l^{2(s+1)} X^{(2)}(y_r)$ where $X^{(2)}$ is of degree 
$2(\kappa-s-1)$. 
Therefore, we showed that for large $l$ there appears two separate
curves, $X^{(1)}(y_p)=0$ and $X^{(2)}(y_r)=0$. This is what is expected
as a property of the algebraic curves probing the instantons: when
sets of instantons are far away from each other then the curve
splits. 

It may be interesting if one could show actual factorization of the
algebraic curve in the limit. One may say that (\ref{factori}) itself is
already the ``factorization.'' The most natural 
polynomial which satisfies (\ref{factori}) in the limit $l\to\infty$
(and similar expression for $X^{(2)}$) 
may be just the product, $X=X^{(1)}(y_p)X^{(2)}(y_r)$, 
but note that the degree of
this product is $2\kappa-2$ which is not what one expects.\footnote{
However, this kind of factorization can
be proven for simple polynomials. For example, suppose we have a
polynomial $X(t)$ with degree 2, and two constants $t_1$ and
$t_2$ with $t_1-t_2\equiv l$ large. And suppose that the polynomial is
approximated as $X\sim l(t-t_1)$ for $t\sim t_1$  and 
$X\sim -l(t-t_2)$ for $t\sim t_2$. 
Then the only possible such polynomial is $X=(t-t_1)(t-t_2)$.
}

As an extreme example, consider the case $s+1=\kappa$.
We take the limit $|a_{\kappa}-a_p| \sim l$ very large
($p=0,1,\cdots,\kappa-1$) and keep all $s_i$ fixed. 
In this limit, $y_\kappa$ goes to infinity 
when we consider $x$ taking values just around 
$a_{0},\cdots, a_{\kappa-1}$. As before, 
we obtain $X \sim l^2 X^{(1)}(y_p)$ where $p=0,1,\cdots,\kappa-1$.
Note that in this case, the Higgs zero locus $X=0$ does not give
an additional isolated point $x_\mu=a_{\kappa\mu}$. 
When $|y_\kappa|$ is small, the
other $|y_p|\sim l$ $(p=0,1,\cdots,\kappa-1)$ is very large, so all the 
$|y_p|$ dependence drops off in $Z$ and $Q$.
Then $Z$ and $Q$ become a function of only $y_\kappa$. In this limit
in general, vanishing of $X$ does not give any solution.

This appears to give a contradiction since this well-separated
$a_\kappa$ would have been identified with a single well-separated
instanton. But this is not the case. 
Note that the JNR ansatz includes $\kappa+1$ locations ($a_i$,
$i=0,1,\cdots,\kappa$) in the space, hence 
the relation between the $a_{i\mu}$ and the
location of the $\kappa$ instantons remains obscure. 
However, for the 't Hooft
instanton which can be obtained from the JNR ansatz by taking 
$s_0\sim |a_{0\mu}|^2 \to \infty$, this relation between $a_i$ 
$(i=1,2,\cdots,\kappa)$ and the location of the instantons is apparent. 
The limit to the 't Hooft instantons uses the scaling of $s_0$ which we
did not consider in the above, so there is no 
contradiction.\footnote{For $\kappa=1$, the 't Hooft instanton is
equivalent to the JNR instanton. If one considers the 
limit $|a_{1\mu}-a_{0\mu}|^2 \to \infty$ in the notation of the JNR
ansatz, one obtains just a large-size single instanton.}

\section{Conclusion and discussions}
\setcounter{footnote}{0}
\label{section4}

In this paper, we considered the identification between the supertube
cross-section and Higgs zero locus of the dyonic instantons, and then
performed a field theoretical calculation for the supertube angular
momentum (\ref{angL}). We showed 
that as we fixed the electric charge $Q_{e}$ and the instanton number
$\kappa$ of the system, the angular momentum is maximized at the
circular loop (section \ref{maxl}).  
This variational problem is closely related to the one for the supertube
in string theory, where one can show that for fixed D0 and F1 charges,
the angular momentum of the supertube is maximized at the circular
cross-section \cite{Mateos:2001pi}. We have shown in section
\ref{dictionary} that, with our proposed dictionary between the charges 
(\ref{instn}) and (\ref{instn2}), the result of the field theoretical
variational problem 
in section \ref{maxl}, based on the identification of the
cross-section of the supertube and the Higgs zero locus of the dyonic
instanton, can be consistently understood. This in turn, is
a strong support for the identification of the cross-section of the
supertubes and the Higgs zero locus of the dyonic instantons.
As a T-dual version of this identification, in section \ref{supercurve}
we also studied the correspondence between the D-helices/supercurves and
the wavy instanton string solutions in the 1+5 dimensional SU(2)
Yang-Mills theory. The instanton strings can be located
without any ambiguity, and the identification of the location with the 
D-helix/supercurve shape is very natural. In this case, we have given a
proof of the dictionary of the charges as a functional of the shape.

The conjecture on the correspondence between the ADHM data and the real
algebraic curves studied in section \ref{section3} is still preliminary,
in the sense that we need to give the precise definition of the
correspondence. 
Although the rough correspondence on
the degrees of the polynomial and the splitting property has been
clarified, we still need 
to specify which set of polynomials should be treated for giving
the on-to map from the set of instanton moduli space. However, it is 
intriguing that the data of the instantons can be understood in a
graphical manner by the algebraic curves, this has been provided through  
the 
identification of the dyonic instantons and the supertubes ending on
the D4-branes. 
Moreover, our results in section \ref{section3} provided the
quantitative evidence 
that, in the large instanton number limit, we can reproduce algebraic 
curve of arbitrary shape
by the Higgs zero locus, echoing the conjecture made in
\cite{Townsend:2004nc}. 

As a remark, consider the maximum value of the angular momentum
(\ref{angL}). Let us compare it with the value expected from the D-brane
picture. For this purpose, we need the dictionary of the charges in the
two pictures given in section \ref{dictionary}. 
For $\kappa=2$, the dictionary (\ref{instn}) leads in particular to the
following equation
\begin{eqnarray}
 \frac{2\pi}{T_{\rm D2}} Q_{\rm D0} Q_{\rm F1} = 
\frac{Q_{\rm e}}{4\pi(q^3)^2} 
\label{upperl}
\end{eqnarray}
with  $T_{\rm D2}$ the tension of the D2-brane.
In fact the combination appearing in the left hand side of the equation
is the precise upper bound for the angular momentum $L_{\rm D2}$ of the
supertube, as shown in  
\cite{Mateos:2001pi,Bak}.\footnote{Note that we have defined the angular
momentum as (\ref{angL}) which is different from the definition in
\cite{Bak}, by a factor.}  
For a circular shape, we use the data (\ref{cm}) to evaluate the
electric charge as $Q_{\rm e} = 16 \pi^2 R^2 (q^3)^2$, 
with which the upper-bound for the angular momentum (\ref{upperl})
is 
\begin{eqnarray}
 L_{\rm supertube}^{\rm (max)} = 4\pi R^2\ .\label{angularbound} 
\end{eqnarray}
On the other hand, 
what we showed in this paper is that the following upper bound exists
for $\kappa = 2$ instantons
\begin{eqnarray}
L \le 2\pi \frac{\sqrt{13} - 1}{6}\ R^2 
\quad (<  L_{\rm supertube}^{\rm (max)})
\, .
\end{eqnarray}
Clearly it lies below the bound expected from the supertube, 
(\ref{angularbound}). Let us discuss the reason for this.
In \cite{Mateos:2001pi}, the authors showed (\ref{angularbound}) is
satisfied by the circular supertubes with {\it{uniform}} electric
$Q_{\rm F1}$ and magnetic $Q_{\rm D0}$ charge densities. We can
therefore conclude from the general analysis in \cite{Mateos:2001pi}
that the field theory 
configuration considered here should at most correspond to a circular
supertube with {\it{non-uniform}} charge distributions. In fact, the
instanton charge distribution can be found non-uniform along the 
closed curve given by the zero of the Higgs field. This suggests that
the saturation of the bound given by the supertube is possible only 
in the limit of large instanton number.

Let us also comment on the definition of the angular momentum 
of the dyonic instantons, which is generally different from that of
supertube.  
There are in fact two other possible ways to define the angular momentum
of the dyonic 
instantons. One is defined directly by the four dimensional integral
(\ref{AngST}) 
\cite{Kim,Eyras:2000dg}.
The other one is
defined through the ADHM data. One can think of the dyonic instanton as
a time-dependent ADHM data, since the charges and the mass of the dyonic
instanton can be understood as time-dependent solution of a massive
sigma model whose target space is the ADHM moduli space \cite{Tong}. 
This is in accord with the fact that the ADHM data corresponds to the
D0-branes, and the supertube consists of the ``running'' D0-branes.
Obviously, these two definitions differ from our definition of the
angular momentum $L$ (\ref{angL}) which comes from
supertubes.  
For example,
these two other definitions are non-vanishing even for one instanton,
while ours vanishes because the Higgs zero becomes a point.
It is
interesting to see how these three definitions 
are different from one another in more details. 
We expect from the
D-brane perspective, for large
instanton numbers \cite{Townsend:2004nc}
they coincide with one another. 
This is also what one would naturally expect for the identification
between dyonic instantons and supertubes to be unequivocal. 
The T-dual example, the wavy instanton
strings and the D-helices/supercurves, may give some clearer insights,
because for those the dictionary is already proven and the
identification is apparent. 

\acknowledgments 
The authors would like to thank P.~Townsend for the collaborations at 
some stages of this project and 
the numerous enlightening discussions throughout the preparation.
H.Y.C.~would like to thank N.~Manton for reading an earlier draft and 
notifying two references.
K.H.~would also like to thank S.~Ketov, K.~Lee and S.~Terashima 
for valuable comments, and DAMTP members for their hospitality. 
H.Y.C. is supported by St.John's college, Cambridge through a
Benefactors' Scholarship. 
The work of M.E.~is supported by Japan Society for the Promotion 
of Science (JSPS) under the Post-doctoral Research Program. 
K.H.~is partly supported by JSPS and 
the Japan Ministry of Education, Culture, Sports, Science and
Technology. 


\newcommand{\J}[4]{{\it #1} {\bf #2} (#3) #4}
\newcommand{\andJ}[3]{{\bf #1} (#2) #3}
\newcommand{\AP}{Ann.\ Phys.\ (N.Y.)}
\newcommand{\MPL}{Mod.\ Phys.\ Lett.}
\newcommand{\NP}{Nucl.\ Phys.}
\newcommand{\PL}{Phys.\ Lett.}
\newcommand{\PR}{ Phys.\ Rev.}
\newcommand{\PRL}{Phys.\ Rev.\ Lett.}
\newcommand{\PTP}{Prog.\ Theor.\ Phys.}
\newcommand{\hep}[1]{{\tt hep-th/{#1}}}

\end{document}